\documentclass{epl}

\title{Fermions at unitarity and Haldane exclusion statistics}

\author{R.K. Bhaduri\inst{1,2}, M.V.N. Murthy\inst{2} \and M.K. 
Srivastava \inst{3}}

\institute{                    
  \inst{1} Department of Physics and Astronomy, McMaster University,
Hamilton, Canada L8S 4M1\\
\inst{2} The Institute of Mathematical Sciences, Chennai 600113, 
India\\
\inst{3} Institute Instrumentation Centre, Indian Institute of 
Technology, Roorkee 247667, India}


\begin{document}

\maketitle

\begin{abstract}

We consider a gas of neutral fermionic atoms at ultra-low temperatures,
with the attractive interaction tuned to Feshbach resonance. We
calculate, the variation of the
chemical potential and the energy per particle as a function of
temperature by assuming the system to be an ideal gas obeying the
Haldane-Wu fractional exclusion statistics. Our results for the
untrapped gas compare favourably with the recently published Monte Carlo
calculations of two groups. For a harmonically trapped gas, the results
agree with experiment, and also with other published work. 

\end{abstract}

Consider a dilute gas of neutral fermionic atoms at a low temperature. 
In general, the low-energy properties of the gas are determined by the
scattering length $a$, the number density $n$, and the temperature $T$
of the gas ( the effective range $r_0$ is small, so that $r_0/|a|
\rightarrow 0$ as $a$ becomes large ).  When the attractive interaction
between the atoms is increased continuously by magnetic tuning from weak
to strong, the scattering length $a$ goes from a small negative to a
small positive value. In between, there is a zero-energy two-body bound
state, and $|a|$ is infinite. The gas is said to be at unitarity in this
situation, and the length scale $a$ drops out. The behaviour of the gas
is expected to be universal at unitarity~\cite{baker}.  Experimentally,
if the temperature is small enough, a BCS superfluid is observed at the
weak end, and a BEC condensate of dimers at the strong end~\cite{regal}.
This was predicted long back by Leggett~\cite{leggett}, who extended the
BCS formalism in a novel fashion to analyse the physical situation.  The
BCS to BEC transition is found to be smooth, with no discontinuity in
properties across the unitary point. There has been much interest
amongst theorists to calculate the properties of the gas in the unitary
regime ($k_f|a|>>1$), where $k_f=(3\pi^2n)^{1/3}$ is the Fermi wave
number of the noninteracting gas. This is a challenging task, since
there is no small expansion parameter, and a perturbative calculation
cannot be done.  In particular, at $T=0$, the energy per particle of the
gas is calculated to be 
\begin{equation} \frac{E}{N}=\xi~\frac{3}{5}
\frac{\hbar^2 k_f^2}{2M}, 
\label{free} 
\end{equation} 
where $\xi\simeq0.44$~\cite{carlson}. The experimental value is about
0.5, but with large error bars~\cite{bart}. Recently, there have been
two Monte Carlo (MC) finite temperature
calculations~\cite{bulag,burovski} of an untrapped gas at unitarity,
where various thermodynamic properties as a function of temperature have
been computed. For a harmonically trapped gas, there are experimental
results~\cite{kinast}, as well as theoretical calculations~\cite{hu}. 

In the unitary regime, the thermodynamic properties have both bosonic
and fermionic features~\cite{bulag}, and it is natural to ask if in this
regime the quasi-particles obey a statistics which is intermediate
between the two. In this paper, we suggest on general grounds that at
unitarity, so far as average properties of the system are concerned, it
should behave like an ideal gas obeying the generalised exclusion
statistics of Haldane~\cite{haldane}. The definition of the statistical
parameter, denoted by $g(>0)$ in the present paper, is based on the rate
at which the number of available states in a system of fixed size
deceases as more and more particles are added to it. The statistical
parameter $g$ assumes values $0$ and $1$ for bosons and fermions
respectively, because the addition of one particle reduces the number of
available states by $g$.  We first deduce the value of $g$ for the
unitary gas from theory using Eq.(\ref{free}), fitting $\xi=0.44$. The 
value of $g$ thus
determined remains the same independent of the nature of confinement as
it should since the microscopic origin of the value of $g$ depends only
on the interaction between fermions and not on how the system
is prepared.  The application of the finite temperature distribution
function~\cite{wu} (called Haldane-Wu statistics in this paper) then
enables us to calculate the temperature dependence of the energy per
particle, and the chemical potential of the unitary gas. Our results for
both trapped and untrapped gases are in good agreement with experiment,
and MC calculations. 

We shall now give the rationale for using Haldane-Wu statistics at
unitarity: 

\begin{itemize}

\item A strong hint that this may be the case comes from the observation
that the kinetic and potential energies scale the same way when there is
no length scale left from the interaction. As is well known, Haldane-Wu
statistics is realised by the Calogero-Sutherland model in one dimension
\cite{murthy}. The potential and kinetic energy both scale the same way
in this model, and both the energy densities scale as $n^3$. Similarly,
fermions in two dimensions interacting with a zero-range potential have
their kinetic and potential energy densities scale as $n^2$, obeying
Haldane-Wu statistics~\cite{rajat}. 

\item In the present case, a compelling evidence comes from
the fact that the second virial coefficient of the gas at unitarity is
temperature independent~\cite{mueller}.  In exclusion statistics, the
scale-invariant interaction between atoms alters the ideal Fermi (Bose)
values of the (exchange) second virial coefficient $+(-) 2^{-5/2}$ by
adding an interacting part~\cite{shankar}. 

\end{itemize}

The above arguments are heuristic and indicative. A quantitative 
understanding can be obtained only when the effective interaction is 
known fully. In the absence of such a theory, in this paper we pursue a 
phenomenological approach where we assume the validity of exclusion 
statistics on the average for quasi-particles which are otherwise 
non-interacting. The effect of interaction is entirely subsumed in 
defining the statistics of the quasi-particles. We first estimate the 
value of the statistical parameter $g$ from the following 
considerations: 

For Haldane-Wu statistics, the distribution
function (or occupancy factor) in a single particle state with energy
$\epsilon_i$ is given by $f_i=(w_i+g)^{-1}$, where $w_i$ obeys the
relation
\begin{equation}
w_i^g(1+w_i)^{1-g}=exp[(\epsilon_i-\mu)\beta]~,
\end{equation}
where $\beta=1/T$, $T$ being the temperature in units of the Boltzmann 
constant. Note from the above that for $g=0$ and $1$, the distribution
function $f_i$ reduces to the familiar bosonic and fermionic forms. It
is also clear that for $T=0$, the occupancy factor is
\begin{eqnarray} 
f_i(T=0)&=&\frac{1}{g}~,~~\epsilon_i <\mu, \nonumber\\
f_i(T=0)&=&0~,~~\epsilon_i>\mu~.
\end{eqnarray}
Now consider $N$ fermionic atoms obeying this statistics at
$T=0$ in a large volume $V$. The new Fermi momentum $\tilde{k_f}$ is
determined from the relation 
$$N=V \frac{1}{g} \frac{2}{(2\pi)^3}\int_0^{\tilde{k_f}} 4\pi k^2 dk~,$$
where we have included a spin degeneracy factor of 2. 
The modified Fermi
momentum $\tilde{k_f}$, from above. is $\tilde{k_f}=g^{1/3} k_f$, 
where $k_f$ is
the fermi momentum of the noninteracting Fermi gas.  It also follows
that the energy per particle of the unitary gas is given by
$$\frac{E}{N}=g^{2/3}~\frac{3}{5}\frac{\hbar^2 k_f^2}{2M}~.$$
Comparing with Eq.(\ref{free}), we see that $\xi=g^{2/3}$, and choosing 
$g=0.29$ gives the generally accepted value of $\xi=0.44$. This therefore 
fixes the only free parameter in the model, namely, $g$ and it should be 
valid independent of temperature and the nature of confinement as it is 
the parameter which determines the statistics of quasi-particles.  

\begin{figure}
\onefigure{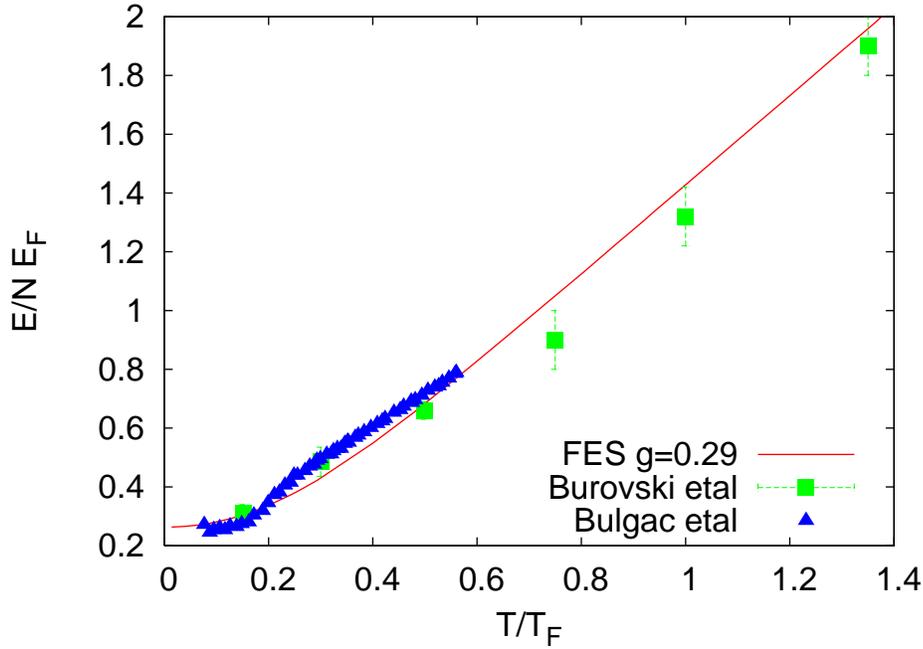}
\caption{Plot of the energy per particle as a function of temperature.
Both the abscissa and ordinate are in units of the free Fermi energy.
The solid line corresponds to our calculations with $g=0.29$. The solid
squares (green) with error bars are the MC calculations
of~\cite{burovski}, and the triangles (blue) are the MC calculations
of~\cite{bulag}. }
\label{fig1} 
\end{figure}

\begin{figure}
\onefigure{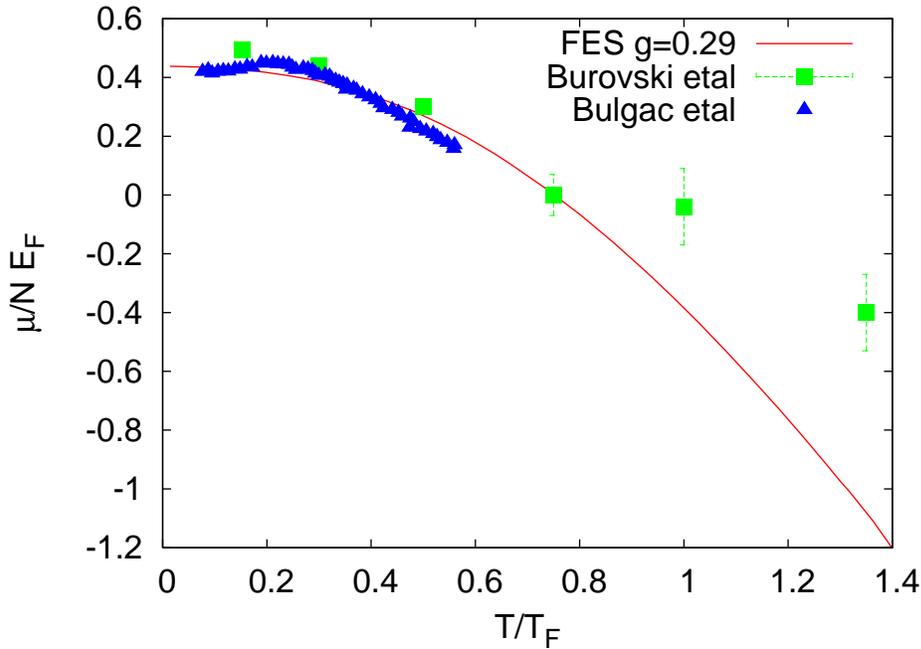}
\caption{Same as in Fig.(\ref{fig1}) for chemical potential plotted as a 
function of temperature.} 
\label{fig2} 
\end{figure}

The main advantage of our model, however, is the calculation of the bulk
properties of the gas as a function of the temperature, and this we
proceed to do now. We follow the well known method (see for example the
paper by Aoyama~\cite{japan}) for this purpose. For a given density of
single-particle states $D(\epsilon)$, we have
\begin{equation}
N=\int_0^{\infty} \frac{D(\epsilon) d\epsilon}{(w+g)},~~~   
E=\int_0^{\infty} \frac{\epsilon D(\epsilon) d\epsilon}{(w+g)}.
\label{general}
\end{equation}
For the $3-$dimensional gas, $D(\epsilon)=C\sqrt{\epsilon}$, where the
constant $C =\frac{3}{2} N\epsilon_f^{-3/2}$. Furthermore,
$\epsilon_f=\frac{\hbar^2k_f^2}{2M}$ is the Fermi energy of the
noninteracting Fermi gas. 
Changing the variable from $d\epsilon$ to $dw$, and using the relation
involving $w$'s given above, one gets after some algebra
\begin{equation}
\frac{3}{2}\left(\frac{T}{\epsilon_f}\right)^{3/2}\int_{w_0}^{\infty}
\frac{dw}{w(1+w)}\left[\ln\left\{
\left(\frac{w}{w_0}\right)^g\left(\frac{1+w}{1+w_0}\right)^{1-g}\right\}
\right]^{1/2}
=1,
\label{once}
\end{equation}
\begin{equation}
\frac{E}{N\epsilon_f}=\frac{3}{2}\left(\frac{T}{\epsilon_f}\right)^{5/2}
\int_{w_0}^{\infty}\frac{dw}{w(1+w)}\left[\ln\left\{
\left(\frac{w}{w_0}\right)^g\left(\frac{1+w}{1+w_0}\right)^{1-g}\right\}
\right]^{3/2}.
\label{twice}
\end{equation}
In the above, $w_0$ is the value of $w$ at $\epsilon=0$. For our choice of
$g=0.29$, the Eq. (\ref{once}) is solved at a given 
$(T/\epsilon_f)$ for $w_0$ numerically, and this
$w_0$ is used in Eq.(\ref{twice}) next to obtain $(E/N\epsilon_f)$. From
the definition of $w_0$, it also follows that the chemical potential
$\mu$ at temperature $T$ obeys the relation 
\begin{equation}
\frac{\mu}{\epsilon_f}=-\frac{T}{\epsilon_f} \left[g\ln w_0+(1-g)\ln
  (1+w_0)~\right ].
\label{thrice}
\end{equation}  
Our results for the energy per particle and the chemical potential (in
units of the noninteracting Fermi energy $\epsilon_f$ ) are 
plotted in Fig.\ref{fig1} and Fig.\ref{fig2} respectively.  

Our results are not sensitive to the fine-tuning of the statistical
parameter $g$. In Fig. \ref{fig1}, we also show, for comparison, the
recent MC calculated points of Bulgac {\it et al.}\cite{bulag} and
Burovski {\it et al.}~\cite{burovski}.  It will be seen that the
agreement is very good, although the chemical potential $\mu$ as
calculated by us starts to differ from Burovski {\it et al} result for
$T/\epsilon_f >0.8)$. 

\begin{figure}
\onefigure{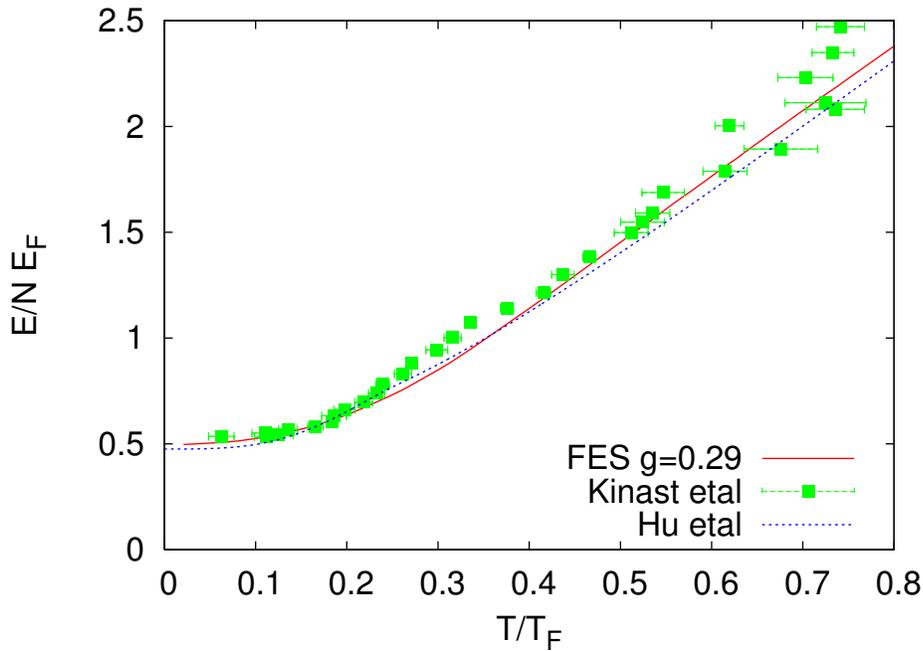}
\caption{Plot of the energy per particle as a function of temperature 
for the confined gas. Both the abscissa and ordinate are  
in units of the free Fermi energy. The solid line corresponds to our 
calculations with $g=0.29$ and the dashed lines corresponds to the 
calculations  presented in Ref.\cite{hu}. The experimental data is taken 
from  Kinast et al \cite{kinast}.
} 
\label{fig3} 
\end{figure}

The finite temperature results are easily generalised for fermions in
harmonic trap. Consider the fermions
at $T=0$. The density of states $D(\epsilon)$, including a spin
degeneracy factor of $2$, is ${\epsilon}^2/(\hbar\omega)^3$, where the 
oscillator parameter is defined as $\omega = (\omega_x 
\omega_y\omega_z)^{1/3}$. It follows
immediately that $\tilde{\epsilon_f}=(3gN)^{1/3} \hbar\omega$, and the energy
$E=g^{1/3} (3N)^{4/3}/4~\hbar\omega$. These results are the same as the
Thomas-Fermi density functional approach of Papenbrock~\cite{pap}.  We
can easily extend these results to finite temperatures using
this density of states in Eq.(\ref{general}). 
\begin{equation}
1= 3\left(\frac{T}{\epsilon_f}\right)^3 
\times \int_{w_0}^{\infty}
\frac{dw}{w(1+w)}\left[\ln\left\{
\left(\frac{w}{w_0}\right)^g\left(\frac{1+w}{1+w_0}\right)^{1-g}\right\}
\right]^2,
\label{oncem}
\end{equation}
\begin{equation}
\frac{E}{N\epsilon_f}=3\left(\frac{T}{\epsilon_f}\right)^4
\times\int_{w_0}^{\infty}\frac{dw}{w(1+w)}\left[\ln\left\{
\left(\frac{w}{w_0}\right)^g\left(\frac{1+w}{1+w_0}\right)^{1-g}\right\}
\right]^3~.
\label{twicem}
\end{equation}
The expression for $\mu$ remains the same as Eq.(\ref{thrice}), although
the numerical values of $w_0$ as a function of $T$ are quite different
from the unconfined gas. We present the results for average energy in 
the trap in Fig.\ref{fig3}, using the same value of
$g=0.29$ since the statistical parameter depends only on the mutual 
interaction and not on the nature of confinement. It will be seen that 
the agreement with experimental data of 
Kinast et al \cite{kinast} as well as the many body calculation of Hu et 
al \cite{hu}  is very good.

Other thermodynamic quantities could be readily calculated.  However we
note that the model cannot yield the two- or many-particle correlation
functions. In this regard, the situation is similar to the
one-dimensional Calogero-Sutherland model~ \cite{csm}, which can be
mapped on to a system of quasi-particles which obey Haldane-Wu
statistics~ \cite{murthy}. But this does not help in obtaining the
correlation functions, for which the full many-body calculation has to
be done. Moreover, the ideal Haldane-Wu gas cannot describe
super-fluidity. Therefore, the main usefulness of the present approach is
its ability to calculate the temperature-dependence of various bulk
properties of a unitary gas with just one free parameter, namely the 
statistical parameter $g$. 

\acknowledgments

We are extremely grateful to Evgeni Burovski, Joaquin Drut and Hui Hu
for sharing the results of their calculations with us.  We would like to
thank Akira Suzuki for useful discussions, and for invaluable help in
preparing the manuscript.  One of the authors (R.K.B.) benefited greatly
from many discussions with Duncan O'Dell.  Thanks are also due to Jules
Carbotte. This research was supported by NSERC of Canada.


\begin{thebibliography}{0}

\bibitem{baker} 
Baker G.A., Phys. Rev. {\bf C60}, 054311 (1999); 
Heiselberg H., Phys. Rev. {\bf A63}, 043606 (2001);
Ho T.-L., Phys. Rev. Lett. {\bf 92}, 090402 (2004).
\bibitem{regal} 
Regal C.A. {\it et al.}, Nature (London) {\bf 424}, 
47 (2003); Zwierlein M.W. {\it et al.}, Phys. Rev. Lett. {\bf91}, 
250401 (2003); Regal C.A. {\it et al.}, Phys. Rev. Lett. {\bf 92}, 
040403 (2004); Zwierlein M.W. {\it et al.}, Nature (London) {\bf 435}, 1046
(12005); Partridge G.B.{\it et al.}, Science {\bf 311}, 503 
(2006).
\bibitem{leggett} 
Leggett A.J., in {\it Modern Trends in the Theory 
of Condensed Matter}, Springer-Verlag Lecture Notes, Vol. 115, edited by 
Peklaski A and Przystawa J.,(Springer-Verlag, Berlin, 1980), p.13
\bibitem{carlson} 
Carlson J., Chang S.-Y., Pandharipande V.R., and
Schmidt K.E., Phys. Rev. Lett. {\bf 91}, 050401 (2003);
Perali A., Pieri P., and Strinati G.C., Phys. Rev. Lett. {\bf 93}, 
100404 (2004).
\bibitem{bart} 
Bartenstein M. {\it et al.}, Phys. Rev. Lett. {\bf 
92},   120401 (2004); Bourdel T. {\it et al.}, Phys. Rev. Lett. {\bf 
93}, 050401   (2004).
\bibitem{bulag} 
Bulgac A., Drut J.E., and Magierski P.,
Phys. Rev. Lett. {\bf 96}, 090404 (2006). 
\bibitem{burovski} 
Burovski E., Prokof'ev N.,
Svistunov B., and Troyer M., Phys. Rev. Lett. {\bf 96}, 160402 (2006).
\bibitem{kinast} 
Kinast J., Turlapov A., Thomas J.E., Qijin Chen, Jelena Stajic and 
Levin K.,
Science {\bf 307} 1296   (2005). 
\bibitem{hu} 
Hu H.. Xia-Ji Lu and Drummond D., Phys. Rev. {\bf A73},
023617 (2006).
\bibitem{haldane} 
Haldane F.D.M., Phys. Rev. Lett. {\bf 67}, 937 (1991).
\bibitem{wu} 
Dasnieres de Veigy A., and Ouvry S., Phys. Rev. Lett. {\bf 
72}, 600 (1994);
Wu Y.-S., Phys. Rev. Lett. {\bf 73}, 922 (1994); 
Isakov S.B., Phys. Rev. Lett. {\bf 73}, 2150 (1994);
Rajagopal A.K.,Phys. Rev. Lett. {\bf 74}, 1048 (1995). 
\bibitem{csm} 
Calogero F., J. Math. Phys. {\bf 10}, 2191 (1969); {\bf
10}, 2197 (1969); Sutherland B., J. Math. Phys. {\bf 12}. 246 (1971); 
{\bf 12}, 251; Phys. Rev. {\bf A4}, 2019 (1971).
\bibitem{murthy} 
Ha Z.N.C., Phys. Rev. Lett.{\bf 73}, 1574 (1994);
Isakov S.B., Phys. Rev. Lett. {\bf 73}, 2150 (1994); 
Murthy M.V.N., and Shankar R., Phys. Rev. Lett. {\bf 73}, 3331 (1994).
\bibitem{rajat} 
Bhaduri R.K., Murthy M.V.N., and Srivastava M.K.,
Phys. Rev. Lett. {\bf 76}, 165 (1996); Srivastava M.K.,
Bhaduri R.K., Law J., and Murthy M.V.N., Can. J. Phys., {\bf 78}9 
(2000).

\bibitem{mueller} 
Ho T.-L., and Mueller E.J., Phys. Rev. Lett. {\bf 92}, 160404 (2004). 
\bibitem{shankar} 
Murthy M.V.N., and Shankar R., Phys. Rev. Lett. {\bf 72}, 3629 (1994). 
\bibitem{japan} 
Aoyama T., Eur. Phys. J. {\bf B 20}, 123 (2001);
cond-mat/0005336 v2
\bibitem{pap} 
Papenbrock T., Phys. Rev. {\bf A72}, 041603 (2005).

\end{thebibliography}
\end{document}